# THE MAGNETIC CURRENT COMPONENT IN
# THE CONDUCTIVITY TENSOR

P. M. Mednis

Novosibirsk State Pedagogical University, Chair of the General and Theoretical Physics, Russia, 630126, Novosibirsk, Viljujsky, 28
e-mail: pmednis@inbox.ru

**Abstract**

The theory of electromagnetic in nature new component of electrical current is suggested. In classical physics approximations for the cases of the free electron plasmas in semi-conductive media, the atom or molecular electrons of liquids and amorphous or polycrystalline solid states medias, tensors of conductivity are calculated. It is shown that the resolution of the additional members of the tensors of conductivity in degrees of the spatial dispersion parameter the series begins minus two degree this parameter, not the only conventional null degree in beginning takes place. The joined optical spatial and time dispersion effects are to be needed for the experimental observation this non-conventional component of the electrical current. Some discussion of this problem is given. In particular, the two-wave propagation effect predicted is discussed.

## 1. Introduction

The microscopic classical Maxwell and Lorentz equations of electrodynamics, such as below in equations (16)-(17) have been written, are asymmetric in form [1]. It is well known that in 1931 P. Dirac supposing the existence of the magnetic charge known as monopole now added to the right sides of these equations the densities of the magnetic charge and current so the equation of electrodynamics will be symmetric. Strangely but nowadays the magnetic charges and the corresponding magnetic currents are not found yet.

Obviously by means of the definition the sum vectors $e_{\pm} = \frac{1}{2}(\mathbf{e} \pm \mathbf{b})$ the Maxwell and Lorentz equations may be written in symmetric form. The symmetric form of the equation (18) realized in this case also gives the physical sense of the vectors $e_{\pm}$. By the way of the historical development of the electrodynamics science this skill symmetry does not resign the Dirac's idea at all. Moreover the reverse statement is incorrect. Indeed, by means of the linear transformation it is impossible to exclude from the Dirac's equations the magnetic charge and current even with aiming to get the asymmetric form of his equations. But this transformation may be useful in the further applications. It is the case at the present work.

So, may it be that the magnetic charge is equal to zero, but the magnetic current as the changing magnetic dipole moments or the changing magnetic dipole moments per unit volume of the charge in a condensed matter really exists? This idea with all demands of relativistic theory was realized early in the work [2]. The difficulty aroused than was the complexness of the non-local in nature new term in the equation of motion of the charged particle. Owing to the proper curl character of the magnetic current considered we transform here all equations to the usual asymmetric form. As a result of this transformation instead of the magnetic current we are dealing now with the new component in the densities of electric charge and current generated by the magnetic current. The formal transformations are considered in the next constructive review section of the present work.

Physical applications of this new component of the electric current are considered in the 3-d section in case of the linear electrodynamics. They are found to become available in a group of the spatial dispersion non gyro-tropic or gyro-tropic effects of electromagnetic wave propagation considered in some details in the classical physics approximation for the cases of the free electron plasmas in semi-conductive matter, the atom and molecular electrons of a liquids and amorphous or polycrystalline medias. For these medias and for the regular atom position in solid states the tensors of the conductivity are calculated.

## 2. The constructive review

In the classical electrodynamics of a condensed matter Maxwell's equations for the *electric field strength vector* **E** and for the *magnetic induction vector* **B** are written as following [3]

$$\nabla \cdot \mathbf{E} = 4\pi(\rho + \rho_0), \qquad \nabla \times \mathbf{E} = -\frac{1}{c}\frac{\partial \mathbf{B}}{\partial t}, \qquad (1)$$

$$\nabla \times \mathbf{B} = \frac{1}{c}\frac{\partial \mathbf{E}}{\partial t} + \frac{4\pi}{c}(\mathbf{j} + \mathbf{j}_0), \qquad \nabla \cdot \mathbf{B} = 0. \qquad (2)$$

Here $\rho_0$ and $\mathbf{j}_0$ are the *volume electric charge density* and the *electric current density vector of* external sources which are introduced now and further on in the course of some arbitrary but under the physical conditions. The densities $\rho$ and $\mathbf{j}$ are induced ones in the matter in the course of the fields action on the charged particles. The $c$ is the velocity of light in vacuum. The $\rho$ and $\mathbf{j}$ are obeying the charge conservation law

$$\nabla \cdot \mathbf{j} + \frac{\partial \rho}{\partial t} = 0 . \tag{3}$$

From the first equation (2) we can find Maxwell's density of the displacement electric current $\frac{1}{4\pi}\frac{\partial \mathbf{E}}{\partial t}$ in vacuum and the usual densities of the electric current $\mathbf{j}$ and $\mathbf{j}_0$ mentioned above. The vector $\mathbf{j}$ may be presented macroscopically in the form [3]

$$\mathbf{j} = \frac{\partial \mathbf{P}}{\partial t} + c \nabla \times \mathbf{M} . \tag{4}$$

Also macroscopically the charge density $\rho$ can be written as
$$\rho = -\nabla \cdot \mathbf{P} . \tag{5}$$

Vectors $\mathbf{P}$ and $\mathbf{M}$ are known as the *polarization* and *magnetization vectors*. The term $\frac{\partial \mathbf{P}}{\partial t}$ is the density of the polarization of a current arising from the time dependence of polarization $\mathbf{P}$ of a matter. This term may include in itself the constant conductivity current in usual sense the electronic or the holes conductivity also. The second term in equation (4) is the curling part of $\mathbf{j}$ arising from the presence of magnetization $\mathbf{M}$ of a matter.

By means of the definition of the *electrical induction vector*

$$\mathbf{D} = \mathbf{E} + 4\pi\mathbf{P} \tag{6}$$

the displacement density of the electric current in a matter is defined in unite form of the sum as follows $\frac{1}{4\pi}\frac{\partial \mathbf{D}}{\partial t}$. By means of the definition of the *magnetic field strength vector* $\mathbf{H}$

$$\mathbf{H} = \mathbf{B} - 4\pi\mathbf{M} \tag{7}$$

similar to equation (6) the equation (1) and (2) may be rewritten in the new widely used form in various practical applications

$$\nabla \cdot \mathbf{D} = 4\pi\rho_0, \qquad \nabla \times \mathbf{E} = -\frac{1}{c}\frac{\partial \mathbf{B}}{\partial t}, \tag{8}$$

$$\nabla \times \mathbf{H} = \frac{1}{c}\frac{\partial \mathbf{D}}{\partial t} + \frac{4\pi}{c}\mathbf{j}_0, \quad \nabla \cdot \mathbf{B} = 0. \tag{9}$$

In progress from these equations to the final results in questions one needs in the addition so called material relationships connecting the inductions vectors of both kinds, $\mathbf{D}$ and $\mathbf{B}$, with the respective *strength vectors* $\mathbf{E}$ and $\mathbf{H}$. This question will be discussed below in the more wide aspects in general.

Now it is a matter of the principle what is the primary: the vectors $\mathbf{P}$ and $\mathbf{M}$ are basic ones or the vector $\mathbf{j}$ and charge density $\rho$ are as such in the (4)? We have no restrictions to the first statement to be true. Even more there is good reason to believe the vectors $\mathbf{P}$ and $\mathbf{M}$ are basic ad hoc. Indeed, as it was proved in my works [2] the microscopically defined polarization vector

$$\mathbf{P}_k = -\frac{1}{4\pi}\frac{q_k}{|\mathbf{r}-\mathbf{r}_k|^3}(\mathbf{r}-\mathbf{r}_k) = \frac{1}{4\pi}\nabla\frac{q_k}{|\mathbf{r}-\mathbf{r}_k|} \tag{10}$$

and magnetization vector

$$\mathbf{M}_k = \frac{1}{c}\mathbf{P}_k \times \mathbf{v}_k = \frac{1}{4\pi c}\nabla \times \frac{q_k \mathbf{v}_k}{|\mathbf{r}-\mathbf{r}_k|} \tag{11}$$

give rise to result by means of formulas (4) and (5) like to the microscopic electric current density as follows

$$\mathbf{j}_k = q_k \mathbf{v}_k \delta(\mathbf{r}-\mathbf{r}_k) = c\nabla \times \mathbf{M}_k + \frac{\partial \mathbf{P}_k}{\partial t}, \tag{12}$$

and the volume electric charge density also

$$\rho_k = q_k \delta(\mathbf{r}-\mathbf{r}_k) = -\nabla \cdot \mathbf{P}_k \tag{13}$$

for one particle with the electric charge $q_k$. In addition to the formulas (10) – (13) the vectors $\mathbf{r}$, $\mathbf{r}_k$ are the space and the $k$-h particle radius - vectors and $\mathbf{v}_k$ is the velocity. The $\delta$-function is the 3-dimensional Dirac's delta function. The full microscopic values mentioned for the system of particles are defined by the sums

$$\mathbf{P}^{(\text{mic})} = \sum_k \mathbf{P}_k, \qquad \mathbf{M}^{(\text{mic})} = \sum_k \mathbf{M}_k, \tag{14}$$

$$\mathbf{j}^{(\text{mic})} = \sum_k \mathbf{j}_k, \qquad \rho^{(\text{mic})} = \sum_k \rho_k. \tag{15}$$

It will be noted here that the formulas (10) and (11) are not unambiguous to be held the equations (12) and (13). On account of the conditions $\nabla \times \mathbf{P}_k = 0$ and $\nabla \cdot \mathbf{M}_k = 0$ takes place and the values of vectors $\mathbf{P}_k$ and $\mathbf{M}_k$ must be vanished with the rate of no less than $1/r^2$ if the distance $r \to \infty$ because of the boundary conditions, the uniqueness conditions in (12) and (13) occur.

The microscopic electrodynamics equations for the *electric field strength vector* $\mathbf{e}$ and for the *magnetic induction vector* $\mathbf{b}$ are written usually [1] as follows

$$\nabla \cdot \mathbf{e} = 4\pi \rho^{(\text{mic})}, \qquad \nabla \times \mathbf{e} = -\frac{1}{c}\frac{\partial \mathbf{b}}{\partial t}, \tag{16}$$

$$\nabla \times \mathbf{b} = \frac{1}{c}\frac{\partial \mathbf{e}}{\partial t} + \frac{4\pi}{c}\mathbf{j}^{(\text{mic})}, \qquad \nabla \cdot \mathbf{b} = 0. \tag{17}$$

These equations are followed by the standard equation of the motion the charge acted by Lorentz force

$$\frac{d\boldsymbol{p}_k}{dt} = q_k \mathbf{e}(r_k, t) + \frac{q_k}{c}\mathbf{v}_k \times \mathbf{b}(r_k, t). \tag{18}$$

Using the vectors (10) and (11) we may introduce far from being formally [2] the densities of *magnetic charge* and *magnetic current* by means of the following definitions

$$\rho^{(m)} = \sum_k \rho_k^{(m)} = \nabla \cdot \mathbf{M}^{(\text{mic})}, \qquad \rho_k^{(m)} = \nabla \cdot \mathbf{M}_k, \tag{19}$$

$$\mathbf{j}^{(m)} = \sum_k \mathbf{j}_k^{(m)} = c\nabla \times \mathbf{P}^{(\text{mic})} - \frac{\partial \mathbf{M}^{(\text{mic})}}{\partial t}, \qquad \mathbf{j}_k^{(m)} = c\nabla \times \mathbf{P}_k - \frac{\partial \mathbf{M}_k}{\partial t}. \tag{20}$$

The values obey the own conservation law

$$\nabla \cdot \mathbf{j}_k^{(m)} + \frac{\partial \rho_k^{(m)}}{\partial t} = 0. \tag{21}$$

The analysis mentioned above yields that $\rho_k^{(m)} = \nabla \cdot \mathbf{M}_k = 0$, so the magnetic charge is equal to zero. As for the magnetic current we have here $c\nabla \times \mathbf{P}_k = 0$, but nevertheless $\mathbf{j}_k^{(m)} = -\frac{\partial \mathbf{M}_k}{\partial t} \neq 0$. This means that the full magnetic current $\mathbf{j}^{(m)} = -\frac{\partial \mathbf{M}^{(\text{mic})}}{\partial t}$ may be introduced in the microscopic electrodynamics equations (16) and (17).

This program were realized in the modified form of the fields equations [2]

$$\nabla \cdot \mathbf{e} = 4\pi\rho^{(\text{mic})}, \qquad \nabla \times \mathbf{e} = -\frac{1}{c}\frac{\partial \mathbf{b}}{\partial t} + \frac{4\pi}{c}\mathbf{j}^{(m)}, \qquad (22)$$

$$\nabla \times \mathbf{b} = \frac{1}{c}\frac{\partial \mathbf{e}}{\partial t} + \frac{4\pi}{c}\mathbf{j}^{(\text{mic})}, \quad \nabla \cdot \mathbf{b} = 0. \qquad (23)$$

This modified equations leads to the new non-local equation of motion of *k*-h particle

$$\frac{d\mathbf{p}_k}{dt} = q_k \mathbf{e}(\mathbf{r}_k, t) + \frac{q_k}{c}\mathbf{v}_k \times \mathbf{b}(\mathbf{r}_k, t) + \int \left[\mathbf{P}_k \times \left(\nabla \times \mathbf{e} + \frac{1}{c}\frac{\partial \mathbf{b}}{\partial t}\right)\right] dV. \qquad (24)$$

The relativistic impulses of the particles as usual are given as

$$\mathbf{p}_k = \frac{m_k \mathbf{v}_k}{\sqrt{1 - \mathrm{v}_k^2/c^2}}. \qquad (25)$$

The equation (24) is not a convenient thing. Owing to the second equation in (22) and definitions (10) and (11) we can calculate now the integral in (24) in explicit form. So the equation of the motion we can write in the usual or conventional Lorentz form

$$\frac{d\mathbf{p}_k}{dt} = q_k \mathbf{e}^*(\mathbf{r}_k, t) + \frac{q_k}{c}\mathbf{v}_k \times \mathbf{b}(\mathbf{r}_k, t), \qquad (26)$$

where the effective starred field $\mathbf{e}^*(\mathbf{r}, t)$ is defined as the sum

$$\mathbf{e}^*(\mathbf{r}, t) = \mathbf{e}(\mathbf{r}, t) + \mathbf{e}^{(\text{ind})}(\mathbf{r}, t). \qquad (27)$$

The induction field is defined as

$$\mathbf{e}^{(\text{ind})}(\mathbf{r}, t) = \frac{1}{c^2}\frac{\partial}{\partial t}\sum_l \left[\frac{q_l \mathbf{v}_l}{|\mathbf{r} - \mathbf{r}_l|} - \frac{1}{2}\nabla \frac{q_l(\mathbf{r} - \mathbf{r}_l) \cdot \mathbf{v}_l}{|\mathbf{r} - \mathbf{r}_l|}\right]. \qquad (28)$$

After exception the field $\mathbf{e}(\mathbf{r}, t)$ from the (22) and (23) we get the equations

$$\nabla \cdot \mathbf{e}^* = 4\pi \rho^{*(\text{mic})}, \qquad \nabla \times \mathbf{e}^* = -\frac{1}{c}\frac{\partial \mathbf{b}}{\partial t}, \qquad (29)$$

$$\nabla \times \mathbf{b} = \frac{1}{c}\frac{\partial \mathbf{e}^*}{\partial t} + \frac{4\pi}{c}\mathbf{j}^{*(\text{mic})}, \quad \nabla \cdot \mathbf{b} = 0. \qquad (30)$$

The starred electric current density vector and the volume electric charge density are

$$\mathbf{j}^{*(\text{mic})} = \mathbf{j}^{(\text{mic})} - \frac{1}{4\pi}\frac{\partial \mathbf{e}^{(\text{ind})}}{\partial t}, \qquad (31)$$

$$\rho^{*(\text{mic})} = \rho^{(\text{mic})} + \frac{1}{4\pi}\nabla \cdot \mathbf{e}^{(\text{ind})}. \qquad (32)$$

It is evident that the conservation law takes place to the full also

$$\nabla \cdot \mathbf{j}^{*(\text{mic})} + \frac{\partial \rho^{*(\text{mic})}}{\partial t} = 0. \qquad (33)$$

The nature of this current is rising from the changing of the $\mathbf{e}^{(\text{ind})}(r,t)$, which was born owing to the change of the magnetic current density $\mathbf{j}^{(m)} = -\frac{\partial \mathbf{M}^{(\text{mic})}}{\partial t}$.

So, the equations (26) and (28) – (33) with the definitions of the $\mathbf{j}^{(\text{mic})}$ and $\rho^{(\text{mic})}$ given by the formulas (15) are forming the self-coincide system of equation describing the classical state of interacting of the field and particles. It is evidently that this dynamic state depends of the initial state of the system. For the macroscopic condensed matter systems this state is not fully defined except the equilibrium statistical case. According to Lorentz [3, 4] the macroscopic fields $\mathbf{E}$ and $\mathbf{B}$ are the average by means of any statistical distribution ones. Hence we can write for the fields

$$\mathbf{E} = \langle \mathbf{e}^* \rangle, \qquad \mathbf{B} = \langle \mathbf{b} \rangle, \qquad (34)$$

for the polarization and magnetization vectors

$$\mathbf{P} = \langle \mathbf{P}^{(\text{mic})} \rangle, \quad \mathbf{M} = \langle \mathbf{M}^{(\text{mic})} \rangle, \qquad (35)$$

and so on for an arbitrary vector $\mathbf{A}^{(\text{mic})}$ meaning the following mathematical operations

$$\mathbf{A} = \int \mathbf{A}^{(\text{mic})} dw, \quad \int dw = \int f \, d\Gamma = 1. \qquad (36)$$

In the latter integral the $f$ and $d\Gamma$ are the distribution function and volume elements in the phase space. Note that the in course of averaging, the following properties take place

$$\langle \nabla \times \mathbf{A} \rangle = \nabla \times \langle \mathbf{A} \rangle, \quad \langle \nabla \cdot \mathbf{A} \rangle = \nabla \cdot \langle \mathbf{A} \rangle, \quad \left\langle \frac{\partial \mathbf{A}}{\partial t} \right\rangle = \frac{\partial \langle \mathbf{A} \rangle}{\partial t}. \tag{37}$$

According to this scheme we get from the (29)-(32)

$$\nabla \cdot \mathbf{E} = 4\pi(\rho^* + \rho_0^*), \qquad \nabla \times \mathbf{E} = -\frac{1}{c}\frac{\partial \mathbf{B}}{\partial t}, \tag{38}$$

$$\nabla \times \mathbf{B} = \frac{1}{c}\frac{\partial \mathbf{E}}{\partial t} + \frac{4\pi}{c}(\mathbf{j}^* + \mathbf{j}_0^*), \quad \nabla \cdot \mathbf{B} = 0. \tag{39}$$

$$\mathbf{j}^* = \langle \mathbf{j}^{*(\text{mic})} \rangle = \langle \mathbf{j}^{(\text{mic})} \rangle - \frac{1}{4\pi}\frac{\partial \langle \mathbf{e}^{(\text{ind})} \rangle}{\partial t}, \tag{40}$$

$$\rho^* = \langle \rho^{*(\text{mic})} \rangle = \langle \rho^{(\text{mic})} \rangle + \frac{1}{4\pi}\nabla \cdot \langle \mathbf{e}^{(\text{ind})} \rangle. \tag{41}$$

And, the conservation law takes place:

$$\nabla \cdot \mathbf{j}^* + \frac{\partial \rho^*}{\partial t} = 0. \tag{42}$$

The densities $\rho = \langle \rho^{(\text{mic})} \rangle$ and $\mathbf{j} = \langle \mathbf{j}^{(\text{mic})} \rangle$ are the usual ones used in standard sense as in equation (1) and (2). The new components of the charge and current densities in a matter are defined by means of the average induction electric field

$$\langle \mathbf{e}^{(\text{ind})} \rangle = \frac{1}{c^2}\frac{\partial}{\partial t}\sum_l \left[ \left\langle \frac{q_l \mathbf{v}_l}{|\mathbf{r} - \mathbf{r}_l|} \right\rangle - \frac{1}{2}\nabla \left\langle \frac{q_l(\mathbf{r} - \mathbf{r}_l) \cdot \mathbf{v}_l}{|\mathbf{r} - \mathbf{r}_l|} \right\rangle \right]. \tag{43}$$

The sense of the macroscopic fields we are using is contained in equation of charge motion

$$\frac{d\mathbf{p}_k}{dt} = q_k \mathbf{E}(\mathbf{r}_k, t) + \frac{q_k}{c}\mathbf{v}_k \times \mathbf{B}(\mathbf{r}_k, t). \tag{44}$$

as in (1) and (2). Note, here and so on the volume electric charge density $\rho_0^*$ and the electric current density vector $\mathbf{j}_0^*$ of external sources are introduced in (38) and (39) also.

For those cases where the dividing kinds of currents have no sense or failed we may write another form of equations (38) and (39) so as it was done elsewhere [3-6] by means of the new definition the electrical induction vector

$$\mathbf{D}'(r,t) = \mathbf{E}(r,t) + 4\pi \int_{-\infty}^{t} \mathbf{j}^*(r,t')\, dt' . \tag{45}$$

With the prime electrical induction vector we have the equations in the form

$$\nabla \cdot \mathbf{D}' = 4\pi \rho_0^*, \qquad \nabla \times \mathbf{E} = -\frac{1}{c}\frac{\partial \mathbf{B}}{\partial t}, \tag{46}$$

$$\nabla \times \mathbf{B} = \frac{1}{c}\frac{\partial \mathbf{D}'}{\partial t} + \frac{4\pi}{c}\mathbf{j}_0^*, \qquad \nabla \cdot \mathbf{B} = 0 . \tag{47}$$

The equation stated above must be added by the material relationships expressing a connection of the inductions vectors with the fields. In this question below we restrict self by the concrete simple but practical situation.

To finish this review we state here the boundary conditions for the uniform division surface. Let the unit normal vector $\mathbf{n}$ be directed from the first matter region to the second one. The tangential vector $\boldsymbol{\tau}$, which is parallel to the division surface, so the $\mathbf{n} \cdot \boldsymbol{\tau} = 0$ takes place. The right sides of the equations (38) - (39) and (46) – (47) are matter independent. As shown in most textbooks from the equations it follows the continuous condition for the tangential components of electric field vector and the discontinuity condition for the normal components of magnetic induction vector. From the left sides of the equations cited it follows the discontinuity condition for the tangential components of magnetic field vector and the discontinuity condition for the normal components of the electrical induction vector. Generally the boundary conditions are written as

$$\mathbf{E}_{1\tau} = \mathbf{E}_{2\tau} , \quad [\mathbf{n} \times (\mathbf{B}_2 - \mathbf{B}_1)] = \frac{4\pi}{c}(\mathbf{i}' + \mathbf{i}_0^*) , \tag{48}$$

$$\mathbf{B}_{1n} = \mathbf{B}_{2n}, \qquad \mathbf{D}'_{2n} - \mathbf{D}'_{1n} = 4\pi(\sigma' + \sigma_0^*) . \tag{49}$$

Here the vector $\mathbf{i}_0^*$ and $\sigma_0^*$ are the surface electric current density vector and the surface electric charge density of the external sources. The vector $\mathbf{i}'$ is the induced surface electric current density vector defined by linear integral

$$\mathbf{i}' = -\frac{1}{4\pi}\int_1^2 \frac{\partial \mathbf{D}'}{\partial t} dl . \tag{50}$$

The linear integration here is carried out in an infinitively small depth of a surface layer. Also the induced surface electric charge density is

$$\sigma' = \frac{1}{4\pi} \int_1^2 \text{div}[\mathbf{n} \times (\mathbf{D}' \times \mathbf{n})] dl \,. \tag{51}$$

## 3. The linear electrodynamics

In the uniform isotropic medium for the linear electrodynamics the material relationships are written in the tensor form [3-6] as usual

$$D'_i(\mathbf{r},t) = \int_{-\infty}^{t} dt' \int d\mathbf{r}\, \hat{\varepsilon}_{ij}(t-t', \mathbf{r}-\mathbf{r}')\, E_j(\mathbf{r}',t') \,. \tag{52}$$

Write now the four-dimensional Fourier transformation of the any vector function $\mathbf{F}(\mathbf{r},t)$:

$$\mathbf{F}(\mathbf{r},t) = \int \mathbf{F}(\omega,\mathbf{k})\, e^{i(\mathbf{k}\mathbf{r}-\omega t)}\, d\omega\, d\mathbf{k} \,, \quad \mathbf{F}(\omega,\mathbf{k}) = \frac{1}{(2\pi)^4} \int \mathbf{F}(\mathbf{r},t)\, e^{-i(\mathbf{k}\mathbf{r}-\omega t)}\, d\mathbf{r}\, dt \,. \tag{53}$$

Then one gets from the (52) the connection between the Fourier transformation components of the electric field and the electric induction

$$D'_i(\omega,\mathbf{k}) = \varepsilon_{ij}(\omega,\mathbf{k})\, E_j(\omega,\mathbf{k}) \,. \tag{54}$$

Here the complex dielectric permittivity tensor $\varepsilon_{ij}(\omega,\mathbf{k})$ is

$$\varepsilon_{ij}(\omega,\mathbf{k}) = \int_0^{\infty} d\tau \int d\mathbf{R}\, e^{-i(\mathbf{k}\mathbf{R}-\omega\tau)}\, \hat{\varepsilon}_{ij}(\tau,\mathbf{R}) \,. \tag{55}$$

This tensor defines the all kinds of the physical phenomenon concerning the wave propagation in a medium. The $\omega$ dependence of the tensor $\varepsilon_{ij}(\omega,\mathbf{k})$ is known as defining the frequency dispersion phenomenon. The essential parameter of the frequency dispersion to be observed is the ratio $\omega/\omega_s$. The values $\omega_s$ are the proper frequency or own-values, or the proper reverse relaxation time. The $\mathbf{k}$ dependence of the tensor $\varepsilon_{ij}(\omega,\mathbf{k})$ is known as defining the spatial dispersion phenomenon. The essential parameter here for the spatial dispersion be observed is the ratio $a/\lambda$. The $a$ value is the linear range of molecular action depending on the system under consideration [5,6], and the $\lambda$ value is the length of the wave. The general mathematical properties of this tensor and the associated physical phenomenon are considered in many textbooks cited above.

To find the explicit form of this tensor one needs well before find the density of the electric current and so is in the choice of an electronic model of a matter and the presentation of the electric current as a function of the coordinates and the velocities of the particles in course of classical approach using the general method [5] or another equivalent method [6]. In case of quantum mechanical approach we must define the corresponding operator of the electric current and use the methods mentioned also.

To begin now we focus here on the simple classical model of the dielectric media consisting of the independent in rest neutral atoms containing the only electron. In this case the averaging on the initial states is the same as the averaging on the distribution function of the time, coordinates and impulses [6,7]. Even more in our case it is sufficient to consider the one $k$-h atom distribution function $f(\boldsymbol{p}_k, \boldsymbol{r}_k, \boldsymbol{R}_k, t)$ of the electron obeying the equation

$$\frac{\partial f}{\partial t} + \mathbf{v}_k \cdot \frac{\partial f}{\partial \boldsymbol{r}_k} - e\left\{\mathbf{E}_a + \mathbf{E} + \frac{1}{c}\mathbf{v}_k \times \mathbf{B}\right\} \cdot \frac{\partial f}{\partial \boldsymbol{p}_k} = -\frac{f - f_{eq}}{\tau}. \quad (56)$$

In this equation $e$ and $m$ are the electron charge module and the rest mass, the impulses is $\boldsymbol{p}_k = m\mathbf{v}_k$, the relaxation time $\tau$ is an abstract parameter now surely to be discussed in concrete situation. The field $\mathbf{E}_a$ is the atomic field acting on the electron in $k$-h atom. The term $-e(\mathbf{E} + \frac{1}{c}\mathbf{v}_k \times \mathbf{B})$ is the Lorentz force acting on the electron at the point $\boldsymbol{r}_k$. The equilibrium distribution function $f_{eq}$ is the Maxwell – Boltzmann function for the electrons and any function $F(\boldsymbol{R}_k)$ independently characterizing the spatial distribution of the centers of atoms in the media

$$f_{eq}(\boldsymbol{p}_k, \boldsymbol{r}_k, \boldsymbol{R}_k) = A\exp\left[-\frac{\boldsymbol{p}_k^2/2m + U(\boldsymbol{r}_k - \boldsymbol{R}_k)}{k_B T}\right] \cdot F(\boldsymbol{R}_k) , \quad \int f_{eq} d\boldsymbol{p}_k d\boldsymbol{r}_k d\boldsymbol{R}_k = 1 . \quad (57)$$

Note the normalization condition here is the 9-dimensional integral. The dynamics of the system is directed by the Hamiltonian function $H_k = p_k^2/2m + U(\mathbf{r}_k - \mathbf{R}_k)$. This is the non-relativistic electron energy in the $k$-h atom. In the first part of (57) $k_B$ is Boltzmann constant and $T$ is absolute temperature. The starred densities (40) and (41) are written now as

$$\mathbf{j}^* = -e\sum_k \int \mathbf{v}_k \delta(\mathbf{r}-\mathbf{r}_k) f(\mathbf{p}_k,\mathbf{r}_k,\mathbf{R}_k,t)\, d\mathbf{p}_k d\mathbf{r}_k d\mathbf{R}_k +$$
$$+ \sum_k \frac{e}{4\pi c^2} \int \left[\frac{\mathbf{v}_k}{|\mathbf{r}-\mathbf{r}_k|} - \frac{1}{2}\nabla \frac{(\mathbf{r}-\mathbf{r}_k)\cdot \mathbf{v}_k}{|\mathbf{r}-\mathbf{r}_k|}\right] \frac{\partial^2}{\partial t^2} f(\mathbf{p}_k,\mathbf{r}_k,\mathbf{R}_k,t)\, d\mathbf{p}_k d\mathbf{r}_k d\mathbf{R}_k, \quad (58)$$

$$\rho^* = -e\sum_k \int [\delta(\mathbf{r}-\mathbf{r}_k) - \delta(\mathbf{r}-\mathbf{R}_k)] f(\mathbf{p}_k,\mathbf{r}_k,\mathbf{R}_k,t)\, d\mathbf{p}_k d\mathbf{r}_k d\mathbf{R}_k +$$
$$- \sum_k \frac{e}{4\pi c^2} \int \nabla\cdot\left[\frac{\mathbf{v}_k}{|\mathbf{r}-\mathbf{r}_k|} - \frac{1}{2}\nabla \frac{(\mathbf{r}-\mathbf{r}_k)\cdot \mathbf{v}_k}{|\mathbf{r}-\mathbf{r}_k|}\right] \frac{\partial}{\partial t} f(\mathbf{p}_k,\mathbf{r}_k,\mathbf{R}_k,t)\, d\mathbf{p}_k d\mathbf{r}_k d\mathbf{R}_k. \quad (59)$$

Also, the time dependent distribution function obey the condition

$$\int f\, d\mathbf{p}_k d\mathbf{r}_k d\mathbf{R}_k = 1. \quad (60)$$

To the first order in the fields the distribution function

$$f = f_{eq} + f^{(1)} \quad (61)$$

obey the equation

$$\frac{\partial f^{(1)}}{\partial t} + \mathbf{v}_k \cdot \frac{\partial f^{(1)}}{\partial \mathbf{r}_k} - e\mathbf{E}_a \cdot \frac{\partial f^{(1)}}{\partial \mathbf{p}_k} - e\left\{\mathbf{E} + \frac{1}{c}\mathbf{v}_k \times \mathbf{B}\right\} \cdot \frac{\partial f_{eq}}{\partial \mathbf{p}_k} = -\frac{f^{(1)}}{\tau}. \quad (62)$$

Because of special form of (57) the term of $\frac{1}{c}(\mathbf{v}_k \times \mathbf{B}) \cdot \frac{\partial f_{eq}}{\partial \mathbf{p}_k}$ vanish. The atomic field $\mathbf{E}_a$ is defined by the $\mathbf{r}_k$ derivatives of electron potential energy

$$e\mathbf{E}_a = \frac{\partial}{\partial \mathbf{r}_k} U(\mathbf{r}_k - \mathbf{R}_k). \quad (63)$$

The solution obeying the condition $f^{(1)} = 0$ at time $t = -\infty$ corresponding the adiabatic switching of the interaction of the field with the atom at the infinite past is

$$f^{(1)}(\mathbf{p}_k,\mathbf{r}_k,\mathbf{R}_k,t) = e\int_{-\infty}^{t} e^{-\frac{t-t'}{\tau}} e^{L(t-t')} \mathbf{E}(\mathbf{r}_k,t') \cdot \frac{\partial f_{eq}(\mathbf{p}_k,\mathbf{r}_k,\mathbf{R}_k)}{\partial \mathbf{p}_k}\, dt'. \quad (64)$$

Here we called into action the Liouville evolution operator [7,8]

$$L = -\mathbf{v}_k \cdot \frac{\partial}{\partial \mathbf{r}_k} + e\mathbf{E}_a \cdot \frac{\partial}{\partial \mathbf{p}_k} \tag{65}$$

acting on any function of the variables $\mathbf{r}_k$ and $\mathbf{p}_k$ standing right side. This operator is corresponding the characteristic equations of the (62)

$$\frac{d\mathbf{p}_k}{dt} = -e\mathbf{E}_a, \quad \frac{d\mathbf{r}_k}{dt} = \mathbf{v}_k. \tag{66}$$

The solution (64) permits us in principle to find the density of the current (58) expressed in terms of the Fourier components of the electric field $\mathbf{E}$. Below we will be interested in two cases. For the free electron case the atomic field $\mathbf{E}_a$ is equal to zero and the action of the operator (65) resulting the shift in the $\mathbf{r}_k$ variables of the field $\mathbf{E}$. So, from the (64) follows the well known solution [6, 7]

$$f^{(1)}(\mathbf{p}_k, \mathbf{r}_k, \mathbf{R}_k, t) = e \int_{-\infty}^{t} e^{-\frac{t-t'}{\tau}} \mathbf{E}(\mathbf{r}_k - \mathbf{v}_k(t-t'), t') \cdot \frac{\partial f_{eq}(\mathbf{p}_k, \mathbf{r}_k, \mathbf{R}_k)}{\partial \mathbf{p}_k} dt'. \tag{67}$$

Note the equilibrium function (57) is the $\mathbf{r}_k$ independent in this case.

For more complicated case of the electron oscillator atom [9-11] the Hamiltonian is written as

$$H_k = \frac{\mathbf{p}_k^2}{2m} + \frac{m\omega_0^2 (\mathbf{r}_k - \mathbf{R}_k)^2}{2}. \tag{68}$$

So, the Liouville evolution operator has the form

$$L = -\mathbf{v}_k \cdot \frac{\partial}{\partial \mathbf{r}_k} + m\omega_0^2 (\mathbf{r}_k - \mathbf{R}_k) \cdot \frac{\partial}{\partial \mathbf{p}_k}. \tag{69}$$

In this case the action of the exponent operator $e^{L(t-t')}$ on any function $F(\mathbf{p}_k, \mathbf{r}_k)$ returns in the form of the shift and the mix of the variables as follows

$$e^{L(t-t')} F(\mathbf{p}_k, \mathbf{r}_k) = F(\tilde{\mathbf{p}}_k, \tilde{\mathbf{r}}_k), \tag{70}$$

where the variables $\tilde{\mathbf{p}}_k$ and $\tilde{\mathbf{r}}_k$ are

$$\tilde{\mathbf{p}}_k = \mathbf{p}_k \cos\omega_0(t-t') + m\omega_0 (\mathbf{r}_k - \mathbf{R}_k)\sin\omega_0(t-t'),$$
$$\tilde{\mathbf{r}}_k = (\mathbf{r}_k - \mathbf{R}_k)\cos\omega_0(t-t') - \frac{\mathbf{p}_k}{m\omega_0}\sin\omega_0(t-t') + \mathbf{R}_k. \tag{71}$$

Taking this fact into account and integrate over time from the (64) we get the distribution function in this case

$$f^{(1)}(\boldsymbol{p}_k,\boldsymbol{r}_k,\boldsymbol{R}_k,t) = e\int_{-\infty}^{t} e^{-\frac{t-t'}{\tau}} \mathbf{E}(\tilde{\boldsymbol{r}}_k,t') \cdot \frac{\partial f_{eq}(\tilde{\boldsymbol{p}}_k,\tilde{\boldsymbol{r}}_k,\boldsymbol{R}_k)}{\partial \tilde{\boldsymbol{p}}_k} dt'. \qquad (72)$$

Our further goal consists of the calculation the electric current density (58) and the estimation the new additional member with the in both cases known ones.

For the free electron case taking into account the Fourier integral in the (53) and the initial distribution of the form

$$f_{eq}(\boldsymbol{p}_k,\boldsymbol{r}_k,\boldsymbol{R}_k) = \frac{1}{V^2}\left(\frac{1}{2\pi m k_B T}\right)^{3/2} \exp\left[-\frac{p_k^2/2m}{k_B T}\right], \quad \int f_{eq} d\boldsymbol{p}_k d\boldsymbol{r}_k d\boldsymbol{R}_k = 1. \qquad (73)$$

we may integrate the (67) to get the result

$$f^{(1)}(\boldsymbol{p}_k,\boldsymbol{r}_k,\boldsymbol{R}_k,t) = -ie\int \frac{e^{i(\boldsymbol{k}\boldsymbol{r}_k-\omega t)}}{\omega - \boldsymbol{k}\mathbf{v}_k + i/\tau} \mathbf{E}(\omega,\boldsymbol{k}) \cdot \mathbf{v}_k \frac{1}{k_B T} f_{eq}(\boldsymbol{p}_k,\boldsymbol{r}_k,\boldsymbol{R}_k) d\omega d\boldsymbol{k}. \qquad (74)$$

The substitution of this function in the (58) with taking into account the conductivity tensor $\sigma_{ij}(\omega,\boldsymbol{k})$ defined by the relationship

$$j_l^* = \int \sigma_{ls}(\omega,\boldsymbol{k}) E_s(\omega,\boldsymbol{k}) e^{i(\boldsymbol{k}\boldsymbol{r}-\omega t)} d\omega d\boldsymbol{k} \qquad (75)$$

leads us to the formula

$$\sigma_{ls}(\omega,\boldsymbol{k}) = ie^2 n \int \left[v_l + \frac{\omega^2}{c^2}\Phi_l\right] \frac{1}{\omega - \boldsymbol{k}\mathbf{v} + i/\tau} \frac{v_s}{k_B T} \left(\frac{1}{2\pi m k_B T}\right)^{3/2} e^{-\frac{p^2}{2mk_B T}} d\boldsymbol{p}. \qquad (76)$$

Here the $\Phi_l$ factor is the $l$ projection of the vector $\boldsymbol{\Phi}$

$$\boldsymbol{\Phi} = \frac{1}{4\pi}\int \left[\frac{\mathbf{v}}{|\boldsymbol{r}-\boldsymbol{r}'|} - \frac{1}{2}\nabla\frac{(\boldsymbol{r}-\boldsymbol{r}')\cdot\mathbf{v}}{|\boldsymbol{r}-\boldsymbol{r}'|}\right] e^{-i\boldsymbol{k}(\boldsymbol{r}-\boldsymbol{r}')} d\boldsymbol{r}' = \frac{\boldsymbol{k}\times\mathbf{v}\times\boldsymbol{k}}{k^4}. \qquad (77)$$

As for the charge, the density of charge (59) we may find by means of the analogical calculations. But it is easily to find it from the (42). It is sufficient to make the only time integration here. Also, integrating by time the density of the current (75) in the (45) one finds easily with the aid of (54) the dielectric permittivity tensor $\varepsilon_{ij}(\omega, \boldsymbol{k})$ in free electron case considered. The latter is needed in researching a wave processes in this media.

From the equation (76) we see that our new component of current is essential when the general condition $\omega \approx kc$ takes place. Some experimental features of this question are considered in the next section.

Consider now the second solution of our problem. For the case of oscillator atom take the initial distribution function normalized exactly the same as in the (73) in the form

$$f_{eq}(\boldsymbol{p}_k, \boldsymbol{r}_k, \boldsymbol{R}_k) = \frac{1}{V} \left( \frac{\omega_0}{2\pi k_B T} \right)^3 \exp\left[ -\frac{p_k^2/2m + m\omega_0^2(\boldsymbol{r}_k - \boldsymbol{R}_k)^2/2}{k_B T} \right]. \tag{78}$$

For the first step we integrate the current density (58) with the aid of (72) and the (53) by the centers of the atoms variables $\boldsymbol{R}_k$. The next step consist the time integration. Note that this integration does not lead to the elementary function in this case. Finally we integrate over the $\boldsymbol{r}_k$ variables. As the result of the operations mentioned we get the following expression for the conductivity tensor

$$\sigma_{ls}(\omega, \boldsymbol{k}) = e^2 n \int d\boldsymbol{p} \left[ \mathrm{v}_l + \frac{\omega^2}{c^2} \Phi_l \right] \int_0^\infty du \left[ \frac{\mathrm{v}_s}{k_B T} \cos\omega_0 u + \frac{k_B T}{m\omega_0} ik_s [\cos\omega_0 u - 1] \sin\omega_0 u \right] \cdot$$

$$\exp\left\{ iu(\omega + i/\tau) - i\boldsymbol{k}\mathbf{v} \frac{\sin\omega_0 u}{\omega_0} - \frac{k_B T [\cos\omega_0 u - 1]^2 k^2}{2m\omega_0^2} \right\} \left( \frac{1}{2\pi m k_B T} \right)^{3/2} e^{-\frac{p^2}{2mk_B T}}. \tag{79}$$

The crushing factor here is the time dispersion internal integral by $u$. Note if $\omega_0 \to 0$, the (79) turns out to be the (76).

If centers of the atoms are fixed in positions characterized by the vectors $\boldsymbol{a}_k$ of their regularity, the initial distribution (78) changes so the reverse volume to be the delta function

$$\frac{1}{V} \to \delta(\boldsymbol{R}_k - \boldsymbol{a}_k). \tag{80}$$

The integration of the current density (58) with the aid of (72) and the (53) by the centers of the atoms' variables $\boldsymbol{R}_k$ is trivial in this case, but this action did not remove the $\boldsymbol{r}_k$ vari-able from the (72). Because of this fact the final result is highly complicated in principle. No more the single factor $v_l + \frac{\omega^2}{c^2}\Phi_l$ appears. Even more the effect of a spatial modulation of the electric field appears. To perform this integration we use the identity

$$e^{-ar^2} = \frac{\sqrt{\pi}}{2a(2\pi)^2}\int e^{-\frac{aq^2}{4}+iqr}dq, \qquad a = \frac{m\omega_0^2}{2k_B T}. \tag{81}$$

As a result for the current density not for the tensor $\varepsilon_{ij}(\omega,k)$ we can write the expression

$$j^* = \frac{e^2}{m}\frac{1}{k_B T}\frac{1}{(2\pi)^3}\frac{1}{(2\pi m k_B T)^{3/2}}\int dp\int dq\int d\omega dk \int_0^\infty du \exp\left\{-\frac{p^2}{2mk_B T}\right\}\exp\left\{-\frac{m\omega_0^2 q^2}{8k_B T}\right\}$$

$$\cdot \exp\{i[[q + k\cos\omega_0 u]\cdot r - \omega t]\}\exp\left\{i\left\{u[\omega + i/\tau] - k\cdot v\frac{\sin\omega_0 u}{\omega_0}\right\}\right\}$$

$$\cdot\left\{(v + v_{eff})\left[E(\omega,k)\cdot\left[(p\cos\omega_0 u + r\, m\omega_0\sin\omega_0 u\,)L(q,k,u) - i\,m\omega_0\sin\omega_0 u\frac{\partial L(q,k,u)}{\partial q}\right]\right]+\right.$$

$$\left. - im\omega_0\sin\omega_0 u\, L(q,k,u)E(\omega,k)\cdot\frac{\partial v_{eff}}{\partial q}\right\}. \tag{82}$$

In this formula the effective velocity is defined as

$$v_{eff} = \frac{\omega^2}{c^2}\frac{(q + k\cos\omega_0 u)\times v \times(q + k\cos\omega_0 u)}{|q + k\cos\omega_0 u|^4}, \tag{83}$$

and the lattice sum factor is

$$L(q,k,u) = \sum_l \exp\{-il\cdot[q + k(\cos\omega_0 u - 1)]\}, \tag{84}$$

where the vectors of a lattice $l$ are

$$l = l_1 a_1 + l_2 a_2 + l_3 a_3. \tag{85}$$

The non com-planar vectors $a_1$, $a_2$ and $a_3$ are the basic vectors of a lattice, the $l_1$, $l_2$ and $l_3$ are integers. As for the conductivity tensor $\varepsilon_{ij}(\omega,k)$, so we must precede the distinguishing procedure of the internal field acting on the electrons. However, this is suffi-ciently difficult separate task, which is far out from the main interests of the present work. Here the main fact is the existence of additional component of the velocity in the (83).

## 4. The normal wave mode peculiarity

The main question is now to make possible the experimental evidence of the existence of the new contribution to the conductivity tensor. So, the formulas (76), (79) and (82) might be used in the normal mode wave propagation process in account of spatial dispersion in various media. In absence of external sources the normal wave solution [5,6] follows from the (45) – (47) like equations if we take the fields in form

$$\mathbf{E} = \mathbf{E}_0 e^{i(\mathbf{kr} - \omega t)}, \quad \mathbf{B} = \mathbf{B}_0 e^{i(\mathbf{kr} - \omega t)}, \quad \mathbf{D}' = \mathbf{D}'_0 e^{i(\mathbf{kr} - \omega t)}. \tag{86}$$

The amplitudes $\mathbf{E}_0$, $\mathbf{B}_0$ and $\mathbf{D}'_0$ are constant. Note the imaginary unit $i$ is denoted here if the $i$ is not the vector or tensor index. The substitution of the (86) at the (46) and (47) gives the equations for the electric field

$$\left( \frac{\omega^2}{c^2} \varepsilon_{ij}(\omega, \mathbf{k}) - k^2 \delta_{ij} + k_i k_j \right) \mathrm{E}_j = 0. \tag{87}$$

The condition of the non-triviality of the solution gives the determinant to be equal to zero

$$\left| \frac{\omega^2}{c^2} \varepsilon_{ij}(\omega, \mathbf{k}) - k^2 \delta_{ij} + k_i k_j \right| = 0. \tag{88}$$

This equation states the connection between the $\omega$ and $\mathbf{k}$ in the form

$$\omega_i = \omega_i(\mathbf{k}), \quad i = 1, 2, \ldots. \tag{89}$$

where the index $i$ numerates different normal modes. The permittivity tensor $\varepsilon_{ij}(\omega, \mathbf{k})$ is expressed from the (45), (54), (75) and the conductivity tensors stated above. So, one finds [6] this connection in the form

$$\varepsilon_{ij}(\omega, \mathbf{k}) = \delta_{ij} + \frac{4\pi i}{\omega} \sigma_{ij}(\omega, \mathbf{k}), \quad \omega \neq 0. \tag{90}$$

Here are the $\delta_{ij} = 1$, if $i = j$, and $\delta_{ij} = 0$, if $i \neq j$, $\delta_{ij}$ is so called Kroneker symbol.

The conventional asymptotic behaviour of the tensor $\varepsilon_{ij}(\omega, k)$ in limit of $k \to 0$ is so $\varepsilon_{ij}(\omega, 0) = \varepsilon_{ij}(\omega)$ [5-7]. If we take into account our new component of the current, so we can see from the (77), that in all three cases considered the $\varepsilon_{ij}(\omega, k \to 0) \propto 1/k^2$. This means that this current component presents no more than highly spatial non-uniform effect. To evolve this current really it is sufficient for our further consideration to keep in mind the resolution of the tensor $\varepsilon_{ij}(\omega, k)$ in the $k$ series up to null degree only. It follows from the equations (76) and (89) that

$$\varepsilon_{ij}(\omega, k) = \delta_{ij} + \frac{4\pi i}{\omega}\sigma_0 \left[\delta_{ij} + \frac{\omega^2}{c^2}\frac{k^2 \delta_{ij} - k_i k_j}{k^4}\right], \qquad \sigma_0 = \frac{ie^2 n}{m(\omega + i/\tau)}. \tag{91}$$

The members of higher order in degrees of $k$ are small by the parameters of $k_B T / mc^2 \ll 1$ and $m\omega_0^2 r_0^2 / mc^2 \ll 1$, where $r_0$ is the atomic radius. Near the resonant frequency the (79) must be used. Instead of the (91) we get now the (91) like equation in which the substitutions of the frequency factor in the $\sigma_0$ were made as follows

$$\frac{1}{\omega + i/\tau} \to \frac{1}{2}\left[\frac{1}{\omega - \omega_0 + i/\tau} + \frac{1}{\omega + \omega_0 + i/\tau}\right]. \tag{92}$$

The formulas derived may be applied to the wave propagation research in semiconductor optic [12] and in the liquids also. In case of crystalline media the (82) must be used. In spite of this formula is aggravated by a hugeness, it contains the (83) factor if $a \ll \lambda$, which leads the same effect in the $k \to 0$ series. It seems the full consideration the question concerning the crystal matter will be made separately in later time.

Let us write the electric field $\mathbf{E}$ be the sum of the longitudinal $\mathbf{E}_\|$ and transverse $\mathbf{E}_\perp$ components relatively to the vector $\mathbf{k}$

$$\mathbf{E} = \mathbf{E}_\| + \mathbf{E}_\perp, \quad \mathbf{E}_\| = E_\| \frac{\mathbf{k}}{k}, \quad \mathbf{E}_\perp \cdot \frac{\mathbf{k}}{k} = 0. \tag{93}$$

For a longitudinal wave the condition $(k^2\delta_{ij} - k_i k_j)E_{\|j} = 0$ is realized. Hence, from (87), (88) and (91) one sees the contribution absence of our current effect in the approximation considered. On the contrary, for a transverse wave component the correspondent dispersion equation yields

$$k^4 - \left(\frac{\omega}{c}\right)^2 \left(1 + \frac{4\pi i \sigma_0}{\omega}\right) k^2 - \left(\frac{\omega}{c}\right)^4 \frac{4\pi i \sigma_0}{\omega} = 0. \tag{94}$$

Define now the refractive index vector **n** from the condition

$$\boldsymbol{k} = \frac{\omega}{c}\boldsymbol{n}. \tag{95}$$

Than one easily finds the refractive index $n = n(\omega)$ solutions

$$n_{1,2}^2 = \frac{\varepsilon}{2} \pm \sqrt{\frac{1}{4}\varepsilon^2 + (\varepsilon - 1)}, \qquad \varepsilon = 1 + \frac{4\pi i \sigma_0}{\omega}. \tag{96}$$

For transparent frequency region, where the conditions $\omega < \omega_0$ and $\omega_0 - \omega \gg 1/\tau$ are fulfilled, the conductivity $\sigma_0$ is imaginary ad hoc leading to $\varepsilon > 1$. So, the two waves propagate in a media, but one of them is non-uniform decaying wave. The first one is the usually observable wave. The latter wave may be observed in the thin isotropic or an-isotropic layers and nearby the boundary of two contacting medias also. It is any surface wave in this case exists taking contribution to the reflected wave. Note, this is not a full internal reflection effect known. If $\varepsilon < 1$ realized, there exist the two waves propagating with different velocities. The degenerate case is realized if square root in (96) is equal to null. If the condition $\varepsilon \gg 1$ takes place the only one wave exists.

The similar two wave solutions as the conventional spatial dispersion effect were considered in [5,6] in limit of a large $\varepsilon$. The estimations were given the waves to be observable experimentally. However, the sense of our solution essentially differs from the known ones. Indeed, taking into account our new component of the electric current, in general case of spatial uniform gyro-tropic media with the weak spatial dispersion effect one can write the phenomenological resolution of the tensor $\varepsilon_{ij}(\omega, \boldsymbol{k})$ in terms of the refractive index vector $\boldsymbol{n}$ components defined in the (95) as follows

$$\varepsilon_{ij}(\omega, \boldsymbol{k}) = \alpha'_{ijlm}(\omega) \frac{n_l n_m}{n^4} + i \gamma'_{ijl}(\omega) \frac{n_l}{n^2} + \varepsilon_{ij}(\omega) + i \gamma_{ijl}(\omega) n_l + \alpha_{ijlm}(\omega) n_l n_m. \quad (97)$$

The first primed tensor terms here are due to the new component of the current stated above. In an isotropic non gyro-tropic media the term proportional to the odd power of $n_l$ vanish. In these media we have the simplified form of the tensor $\varepsilon_{ij}(\omega, \boldsymbol{k})$:

$$\varepsilon_{ij}(\omega, \boldsymbol{k}) = \alpha'_{ijlm}(\omega) \frac{n_l n_m}{n^4} + \varepsilon_{ij}(\omega) + \alpha_{ijlm}(\omega) n_l n_m. \quad (98)$$

In our case we have taken into account the first two members of the (98). On the contrary, in the works [5,6] have been used the last two ones. It is not difficult to show that the (98) may be expressed in term of the five separate parameters $\varepsilon(\omega)$, $\alpha_1(\omega)$, $\alpha'_1(\omega)$, $\alpha_2(\omega)$ and $\alpha'_2(\omega)$, which are to be calculated for any model of the media, in the form

$$\varepsilon_{ij}(\omega, \boldsymbol{k}) = \left[ \varepsilon(\omega) - \alpha_1(\omega) n^2 - \alpha'_1(\omega) \frac{1}{n^2} \right] \delta_{ij} - \alpha_2(\omega) n_i n_j - \alpha'_2(\omega) \frac{n_i n_j}{n^4}. \quad (99)$$

The united dispersion equation for the transverse wave containing the $\alpha'_1(\omega)$ also is

$$n^4 - \frac{\varepsilon(\omega)}{1 + \alpha_1(\omega)} n^2 + \frac{\alpha'_1(\omega)}{1 + \alpha_1(\omega)} = 0. \quad (100)$$

The calculation the tensor components in (97) is the main task of the linear optic of any media. The substitution of these components in (88), with account the (95), leads to the generalized Augustin Fresnel equation for refractive index $n$ as a function of the frequency $\omega$.

## 5. Conclusion

The main result of this paper is the additional electric current component in (39), (40) and (45) really exists. This statement does not contradict the general physical principle demands such as the relativistic theory principles or the causality principle, etc. The first possible experimental observations of these current components are expected to be in the optical region phenomenon. For this reason in simple classical physic approximation the conductivity tensor has been calculated. The transverse optical wave propagation considered. It will be useful for the real experimental detecting of the new current effect.